\documentclass{article}
\usepackage{graphicx}
\usepackage{amstext}
\begin{document}
\twocolumn[
\title{Detailed structure and dynamics in particle-in-cell simulations of the lunar wake}
\author{Paul C. Birch and Sandra C. Chapman}
\date{\today}

\maketitle
]

\begin{abstract}
The solar wind plasma from the Sun interacts with the Moon, generating a wake structure behind it, since the Moon is to a good approximation an insulator, has no intrinsic magnetic field and a very thin atmosphere.
The lunar wake in simplified geometry has been simulated via a
1$\frac{1}{2}$D electromagnetic particle-in-cell code, 
with high resolution in order to resolve the full phase space dynamics of both electrons and ions.
The simulation begins immediately downstream of the moon, before the solar wind has infilled the wake region, then evolves in the solar wind rest frame.
An ambipolar electric field and a potential well are generated by the electrons, which subsequently create a counter-streaming beam distribution, causing a two-stream instability which confines the electrons.
This also creates a number of electron phase space holes.
Ion beams are accelerated into the wake by the ambipolar electric field, generating a two stream distribution with phase space mixing that is strongly influenced by the potentials created by the electron two-stream instability. The simulations compare favourably with WIND observations.
\end{abstract}


\section{Introduction}

A wake structure is formed as the solar wind flows past the Moon.
The Moon is to a good approximation an insulator, has no intrinsic magnetic field and a very thin atmosphere. Measurements by Lunar Prospector \cite{lin98} and magnetohydrodynamic simulations \cite{harnett00} of the magnetic fields on the Moon indicate a typical scale height of 100km for miniature magnetospheres around the magnetic anomalies (these crustal remnant fields will be neglected here). This results in the Moon being a sink for solar wind particles that collide with it. Directly behind the Moon, there will therefore be a void in which no solar wind particles are present. As the solar wind continues beyond the Moon, this wake is filled in.

Explorer 35 and the Apollo sub-satellites made the first measurements of the lunar wake, albeit with very low resolution. They were able however, to detect a significant depletion in density behind the Moon.
The WIND spacecraft used the Moon for a gravitational assist on December 27, 1994, passing at a distance of 6.5 lunar radii ($R_L$), through the lunar wake. 
With all its instruments switched on, a slice of the structure of the lunar wake was obtained. These observations showed significant ion and electron depletions, rarefaction waves travelling away from the wake, increased electron temperatures, constant ion temperatures, counter-streaming ion beams, an increase in magnetic field strength and a number of electromagnetic and electrostatic waves \cite{ogilvie96,bosqued96,owen96,kellogg96,farrell96,farrell97,bale97a,bale97b}.

Analytical descriptions of the processes involved in filling lunar wake have been considered \cite{ogilvie96,farrell97,borisov00}. Ogilvie {\it et al.} \cite{ogilvie96} detected the counter-streaming beam distribution, creating a two-stream instability. Farrell {\it et al.} \cite{farrell97} then analytically considered a two-stream ion distribution with a background of electrons as the process by which the wake is replenished.

Simulations of this nature have previously been undertaken \cite{farrell98,birch01a,birch01b}. Intriguingly, whereas Farrell {\it et al.} \cite{farrell98} find that the ion dynamics dominate the interaction, the higher phase space resolutions of Birch and Chapman \cite{birch01a,birch01b} established that the electron dynamics dominate the evolution of the wake. In this paper we present new results from these simulations, including detailed particle density and electric field descriptions, a comprehensive analysis of the electron instabilities occurring in the wake, and a detailed description of the structure and dynamics of electron phase space holes generated in the wake that are not found in previous simulations.

\section{The Particle-in-Cell Simulation}

The simulation was performed using a
self consistent, fully relativistic, collisionless 1$\frac{1}{2}$D 
particle-in-cell (PIC) code. In these simulations, the ion and electron distribution functions are represented by a
collection of ``superparticles''. These ``superparticles'', along
with the electric and magnetic fields, can exhibit both full ion and electron
kinetics. The velocities of the ``superparticles'' and the fields have vector
components in three dimensions and all quantities depend on one spatial coordinate, Y, and time.
Since $\nabla\cdot\mathbf{B} = 0$, the component of $\mathbf{B}$ along the simulation is constant, but given this restriction, the simulation is fully electromagnetic.
The phase space resolution of the simulation scales as $\sim\sqrt{n}$, where $n$ is the number of particles per cell. As in Birch and Chapman \cite{birch01a}, 
$\sim 2500$ particles per cell were used, compared with $\sim 80$ particles per cell used by Farrell {\it et al.} \cite{farrell98}.

\begin{figure}
\begin{center}
\includegraphics[width=8.5cm]{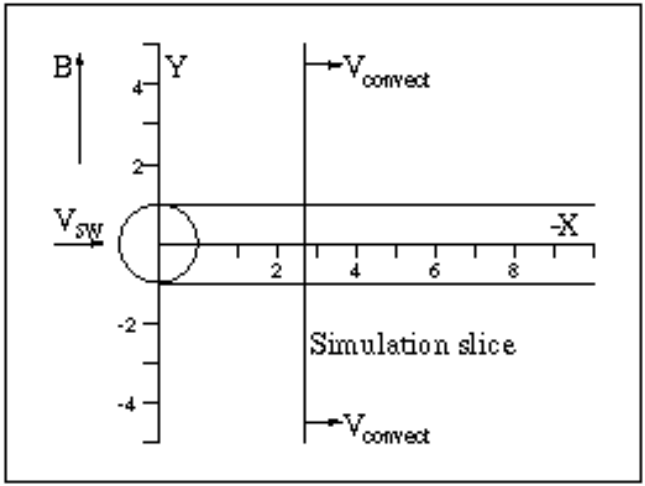}
\caption{Simulation geometry.}
\label{fig:1dsimgeom}
\end{center}
\end{figure}

Figure~\ref{fig:1dsimgeom} shows the geometry of the simulation. 
The simulation box begins directly behind the Moon, perpendicular to the solar wind flow ($-\text{X}$ direction) and parallel to the Interplanetary Magnetic Field (IMF) (Y direction).
A higher dimensional simulation would be required to describe the
lunar wake when the IMF is at oblique angles, and bulk flows
other than in the solar wind flow direction, as the simulation is performed
in the rest frame of the solar wind with zero convection electric field.
The IMF will pass almost unaltered through
the Moon since the Moon can be assumed to be a perfect insulator and
therefore no currents can flow to
change the field. The simulation begins with a void in an otherwise
Maxwellian
distribution of ions and electrons which mimics the removal of solar wind
particles by the Moon.
Full shadowing is appropriate since the electron and ion gyroradii
(1.2km and 47km respectively) are significantly smaller than one lunar radius ($\sim$1738km).
As the simulation box moves with the solar wind, the simulation run time
can be equated to a distance behind the moon, effectively building up a
2D picture of the wake, provided the IMF and solar wind flow remain constant.

An ion to electron mass ratio of 20 has been chosen in order to evolve both ion and electron dynamics. This alters
the simulation's ion plasma frequency and ion thermal velocity to be
\begin{eqnarray}
\omega_{pi} & = & \left( \frac{n_i e^2}{20m_e \varepsilon_0} \right)^{1/2}
= \frac{\omega_{pe}}{\sqrt{20}} \\
v_{thi} & = & \left( \frac{k_B T_i}{20m_e} \right)^{1/2} 
= \frac{v_{the}}{\sqrt{20}}
\end{eqnarray}
since $T_e = T_i$ and $n_e = n_i$. Since the
actual solar wind velocity is 25 times the ion thermal velocity $v_{thi}$, the
simulation also requires this relationship.  Thus, the distance behind the
moon is given by, $\text{X} = V_{conv}t$, where $V_{conv} = 25v_{thi}$ and $t$ is measured in plasma periods, $\omega_{pe}^{-1}$. This is equivalent to 
$\text{X} = 0.274R_Lt$ where $t$ is in plasma periods and $R_L$ is the radius of the moon, i.e. half the width of the void. In reality, $v_{the}>V_{conv}>v_{thi}$; this may affect the magnitude of the potential and effectively rescales $\text{X} = V_{conv}t$.

The simulation run time is limited by the length of the simulation box. If a wake related structure passes over the boundary and interacts with other structures moving in the opposite direction, the simulation must be terminated since this introduces an unphysical distribution. In these simulations a 2048 grid cell simulation box has been used (grid cell $=$ Debye length, $\lambda_D$) which equates to $\pm8R_L$ either side of a $256\lambda_D$ initial void.
An initial void of $128\lambda_D$ has also been simulated to verify that structures found are not a result of the choice of the size of the initial void.

\section{Results of the Simulations}

We begin by discussing the structures
and instabilities found 
and comparing them with theoretical predictions and other simulations. The results will then be compared to the WIND observations at $6.5R_L$.

\subsection{Particle Densities}

\begin{figure*}
\begin{center}
\includegraphics[width=0.815\textwidth]{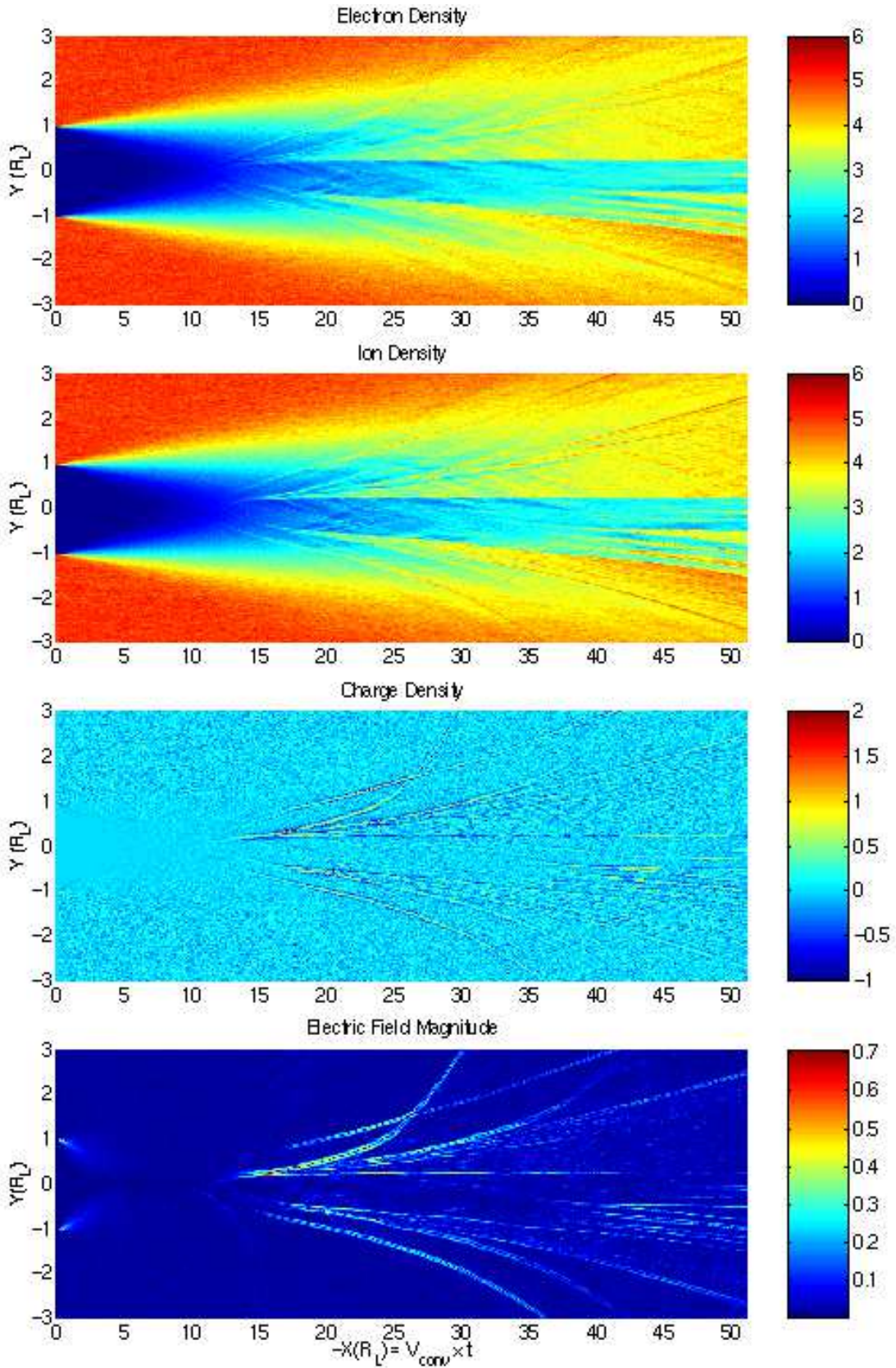}
\caption{Distance along the simulation box Y($R_L$) against time (equated to distance behind the moon) X($R_L$) showing (a) electron density evolution ($\text{cm}^{-3}$); (b) ion density evolution ($\text{cm}^{-3}$); (c) charge density evolution ($e/\text{cm}^{3}$); (d) electric field magnitude evolution ($|\text{E}| / \text{Vm}^{-1}$). The simulation box extends to $\pm8R_L$. The ambient solar wind electron and ion densities are $5\text{cm}^{-3}$.}
\label{fig:big}
\end{center}
\end{figure*}

The density profiles along the simulation box [Y($R_L$)] are plotted on a colour plot at regular intervals [along X($R_L$)], with colour representing density. Figure~\ref{fig:big}(a) shows the evolution of the electron density over the entire simulation.

The first structure to evolve is the rarefaction wave travelling away from the wake at the ion sound speed, $C_S$. This is due to the electrons moving into the wake, and more electrons from the undisturbed plasma filling the space they leave. Ogilvie {\it el al.} \cite{ogilvie96} used analytic solutions of plasma expansion into a vacuum and showed that in such a step in density of a plasma, from $N_0$ to a vacuum at $x=0$ and $t=0$, the expression for particle densities at later times along $x$ are given by the following
\begin{eqnarray}
N_s = N_0 \exp \left[ - \left( \left( \frac{x}{C_S t} \right) + 1 \right) \right]
\end{eqnarray}
where $N_s$ is the density of particle species $s$, $N_0$ is the ambient solar wind density and $C_S$ is the ion sound speed:
\begin{eqnarray}
C_S = \left( \frac{k_B T_e}{m_i} \right)^{1/2}
\end{eqnarray}
where $T_e$ is the electron temperature and $m_i$ is the ion mass. These equations describe a rarefaction wave travelling away from the density step ($x=0$) in the $-x$ direction (away from the wake) at the ion sound speed and an exponential decrease in density in the $+x$ direction (into the wake). In the electron density plot (Fig.~\ref{fig:big}(a)), the rarefaction wave is approximately where the density has dropped to $4\text{cm}^{-3}$.

Initially, the ions behave similarly to the electrons. Figure~\ref{fig:big}(b) shows the evolution of the ion density over the entire simulation. Again, there is a rarefaction wave in the ion density. This wave can be seen to be coincident with the electron rarefaction wave, and therefore also travelling at the ion sound speed in agreement with the above equations. By taking the electron density away from the ion density at all points in Figs.~\ref{fig:big}(a) and (b), a charge density plot can be produced (Fig.~\ref{fig:big}(c)).

Further behind the Moon, once a sufficient number of electrons have entered the wake, more structures begin to appear. After $-\text{X}$($R_L$) $\approx10R_L$ enhancements of the electron density and ion density in the central wake region can be seen. These are due to the electrons and ions streaming into the wake and setting up a counter-streaming beam distribution. Such a distribution is unstable and subsequently a two-stream instability grows. This instability is discussed in more detail in Sec.~III-D. The obvious density enhancements emerging from this region are due to the electrons being trapped by electric potentials generated by the instability. Narrower density enhancements are also generated by the two-stream instability.  In these structures, the electron density forms a double peaked structure about a single peak in the ion density. From Fig.~\ref{fig:big}(c), the charge density plot, it can be seen that this is the main deviation from quasineutrality in the simulations. These are electron phase space holes coupled with propagating ion structures which we will see are ion acoustic solitons and are examined in more detail in Sec.~III-E.

The majority of the structures associated with the lunar wake are confined to within the rarefaction wave. Some electron phase space holes, however, are accelerated above the ion sound speed and overtake the rarefaction wave. The other detectable structures again occur after $-\text{X}$($R_L$) $\approx10R_L$, generally just outside the wake. These are slight fluctuations in the electron density, travelling away from the wake at velocities between $\sim1.8v_{the}$ and $\sim3.4v_{the}$. They are caused by the most energetic electrons which have travelled right across the wake and out on the other side, setting up a bump-on-tail distribution. This distribution is unstable and again, an instability is grown, resulting in these electron density fluctuations. This is an example of the electron bump-on-tail instability and is discussed further in Sec.~III-F.

Comparing these density plots with those produced by the simulations with an initial void $128\lambda_D$ wide shows very little difference. The main difference is the distance between the double electron density peaks. They appear wider than in the $256\lambda_D$ simulations, however, they are actually similar sizes in Debye lengths (not shown here).

The simulations by Farrell {it et al.} \cite{farrell98}, display some similarities with these higher resolution simulations. \linebreak Rarefaction waves and a two-stream instability occur, which shall be discussed further in Sec.~III-D. No electron phase space holes are simulated nor is the bump-on-tail instability detected.

\subsection{Electric Fields and Potentials}

Figure~\ref{fig:big}(d) shows the magnitude of the electric field on the same X, Y scale as Fig.~\ref{fig:big}(c). The first electric field structure to appear is the ambipolar electric field. Significant increases in the electric field magnitude can be seen at Y($R_L$) $=\pm1R_L$. The electric fields at these points are directed into the wake.
Since the electrons have a higher thermal velocity than the ions ($T_e = T_i$ and $m_e < m_i$) they move into the wake first, setting up the electric fields pointing into the wake. These fields then accelerate the ions into the wake. These initial charge separations can also be seen at the beginning of the charge density plot.

\begin{figure}
\begin{center}
\includegraphics[width=8.5cm]{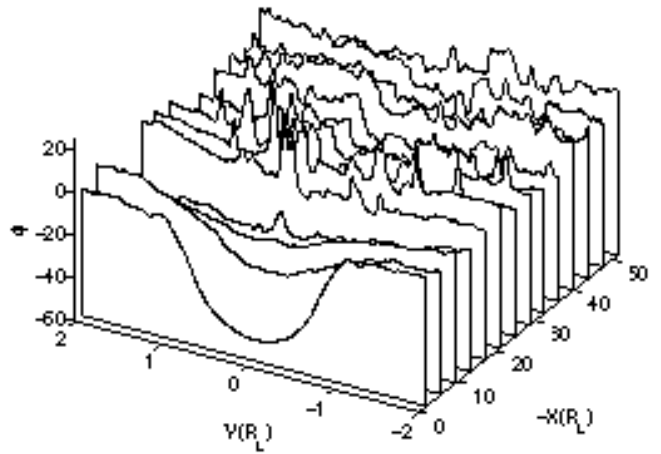}
\caption{Electric potentials ($\text{V}$), beginning at $6.9\omega_{pe}^{-1} \Rightarrow 1.9R_L$, then every $13.8\omega_{pe}^{-1} \Rightarrow 3.8R_L$.}
\label{fig:realepot}
\end{center}
\end{figure}

Figure~\ref{fig:realepot} shows the electric potential calculated at regular intervals from the electric field data. The ambipolar electric field generates the drop in electric potential. This eventually disappears as the ions are accelerated to cancel the charge separation developed by the electrons. The average electric field magnitude decreases back to almost zero, after which the two-stream instability begins to grow.

The two-stream instability produces a number of \linebreak peaks in the electric potential.
These peaks are associated with the bipolar electric field structures, which appear as a double peaked structures in the electric field magnitude plot (Fig.~\ref{fig:big}(d)). Electrons are trapped in these electric potential peaks, oscillating back and forth. Some of these peaks evolve to become electron phase space holes and move out of the wake, coinciding with the double electron density peak about a single ion peak. 

The electric potential peak which grows wider as the simulation evolves, remains nearer the centre of the \linebreak wake. This still has electrons oscillating back and forth as they are confined within the peak. In Fig.~\ref{fig:big}(a), the electron density plot, oscillating density perturbations can be detected in these regions. The electric potential peak becomes wider as more particles move in from outside the wake. It is these structures that will ultimately coalesce to fill the wake.

\subsection{Ion and Electron Phase Space}

\begin{figure}
\begin{center}
\includegraphics[width=8.5cm]{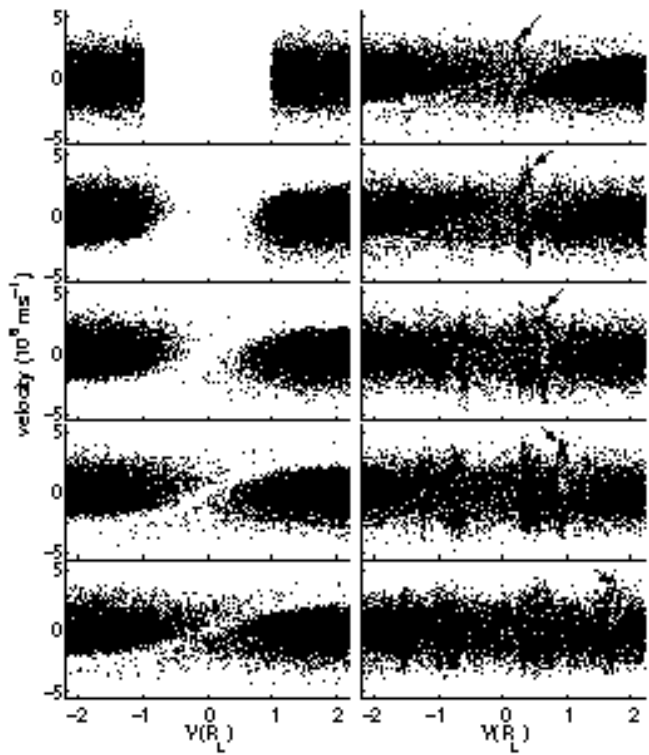}
\caption{Electron phase space evolution [Y($R_L$) against $v_{\text{Y}}$], beginning at $0\omega_{pe}^{-1} \Rightarrow 0R_L$, then every $11.1\omega_{pe}^{-1} \Rightarrow 3.0R_L$.}
\label{fig:realeps}
\end{center}
\end{figure}

\begin{figure}
\begin{center}
\includegraphics[width=8.5cm]{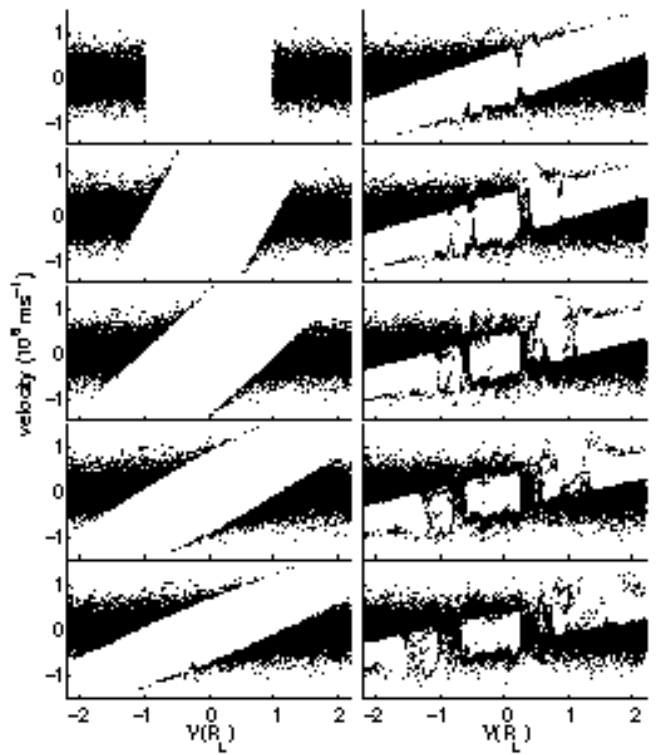}
\caption{Ion phase space evolution [Y($R_L$) against $v_{\text{Y}}$], beginning at $0\omega_{pe}^{-1} \Rightarrow 0R_L$, then every $11.1\omega_{pe}^{-1} \Rightarrow 3.0R_L$.}
\label{fig:realips}
\end{center}
\end{figure}

Figures~\ref{fig:realeps} and~\ref{fig:realips} show the phase spaces (Y, $v_{\text{Y}}$) for the electrons and ions respectively. Note the different velocity scales on both figures; as described earlier, the ion thermal velocity is less than the electron thermal velocity ($v_{the} = \sqrt{20}v_{thi}$).

Initially, the distributions are Maxwellian with a gap at $\text{Y}=[-1,1]R_L$. This void mimics the removal of the solar wind particles by the Moon. In the second phase space plot ($t=11.1\omega_{pe}^{-1}$  $\Rightarrow$ $3.0R_L$), the electrons have begun to move into the wake, followed by the ions. Notice that the ions have been accelerated as they move into the wake. This is due to the ambipolar electric field described earlier. The electrons continue their thermal expansion into the void, setting up counter-streaming electron beams, as seen in the fourth phase space plot ($t=33.3\omega_{pe}^{-1}$ $\Rightarrow$ $9.0R_L$). Now the distribution is unstable and the two-stream instability grows. The first signs of the instability in ion and electron phase space can be seen in the fifth plot ($t=44.4\omega_{pe}^{-1}$ $\Rightarrow$ $12.0R_L$). At $\text{Y} = -0.25R_L$ there is a slight kink in the phase space of the ion beam travelling in the $-\text{Y}$ direction. This is a direct consequence of a peak in the electric potential generated by the two-stream instability. In the next phase space plot ($t=55.5\omega_{pe}^{-1}$ $\Rightarrow$ $15.0R_L$), more disturbances in the phase spaces of the ion beams indicate more disruption by the two-stream instability. Now the disturbances can easily be seen in the equivalent electron phase space plot. There clearly has been a phase space vortex generated at $\text{Y} = +0.25R_L$. This is one of the electron phase space holes and its associated propagating ion structure, seen in the ion phase space as a bump on the otherwise smooth beam. In the electron phase space plots 6 through 10, this and other electron phase space holes can be seen to travel in both directions along Y at various velocities. Their associated ion structures can also be tracked in the same way. One such electron hole has been indicated in Fig.~\ref{fig:realeps}, having carefully followed it with other phase space plots.

\subsection{Two-Stream Instability in the Central Wake Region}

As described in previous sections, the counter-streaming particles set up a two-stream instability, which \linebreak ultimately cause the wake to fill in. In this section, this instability is studied in more detail.

Given a distribution of two electron beams travelling at $\pm V_0$ and two ion beams also travelling at $\pm V_0$, the linear dispersion relation is
\begin{eqnarray}
\frac{\omega_{be}^2}{(\omega - kV_0)^2} &+& \frac{\omega_{be}^2}{(\omega + kV_0)^2} + \frac{\omega_{bi}^2}{(\omega - kV_0)^2} + \nonumber \\ &+& \frac{\omega_{bi}^2}{(\omega + kV_0)^2} = 1
\end{eqnarray}
where $\omega_{be}$ is the electron beam plasma frequency, $\omega_{bi}$ is the ion beam plasma frequency. This gives an expression for the frequency
\begin{eqnarray}
\omega^2 & = & k^2 V_0^2 + (\omega_{be}^2 + \omega_{bi}^2) + \nonumber \\ & \pm & \sqrt{\omega_{be}^2 + \omega_{bi}^2}\sqrt{\omega_{be}^2 + \omega_{bi}^2 + 4 k^2 V_0^2}
\label{eqn:wdr}
\end{eqnarray}
The wavenumber of the fastest growing wave mode of the two-stream instability can be determined by considering
\begin{eqnarray}
\frac{\partial \gamma}{\partial k} = 0 \Rightarrow
k = \frac{\sqrt{3}\sqrt{\omega_{be}^2 + \omega_{bi}^2}}{2 V_0}
\label{eqn:kmax}
\end{eqnarray}
at which the growth rate is
\begin{eqnarray}
\gamma_{max} = \frac{\sqrt{\omega_{be}^2 + \omega_{bi}^2}}{2}
\end{eqnarray}

By estimating the electron and ion densities from the particle density plots (Figs.~\ref{fig:big}(a) and (b)) to be a third of the ambient solar wind density, an estimate for the maximum growth rate can be found
\begin{eqnarray}
\omega_{be} & \approx & \frac{\omega_{pe}}{\sqrt{3}} \\
\omega_{bi} & \approx & \frac{\omega_{pi}}{\sqrt{3}} = \frac{\omega_{pe}}{\sqrt{60}} \\
\gamma_{max} & \approx & \frac{\omega_{pe}}{2} \sqrt{ \frac{1}{3} + \frac{1}{60} } \approx 0.3\omega_{pe}
\end{eqnarray}

\begin{figure}
\begin{center}
\includegraphics[width=8.5cm]{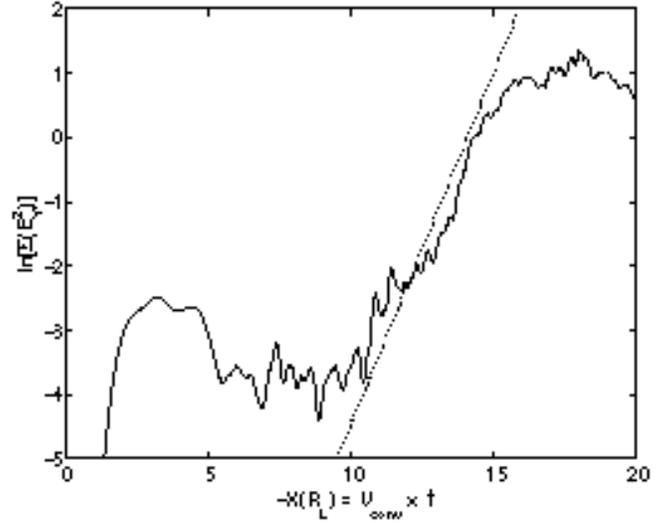}
\caption{$\ln(\sum{E_{\text{Y}}^2})$ plotted against time (equated to distance behind the moon) X($R_L$) for the region of the simulation box $\text{Y} = [-0.5,0.5]R_L$. The dotted line represents the estimated maximum growth rate for this two-stream instability.}
\label{fig:realefmts}
\end{center}
\end{figure}

The growth rate can be measured by summing the $\text{E}^2$ in the region of interest and taking the natural logarithm. Figure~\ref{fig:realefmts} shows the result of this calculation over the first $\sim 73\omega_{pe}^{-1}$ of the simulation. The dotted line shows the estimated maximum growth rate, which follows the increase in $\ln(\sum{{E}^2})$ quite closely. The initial increase in $\ln(\sum{{E}^2})$ is due to the ambipolar electric field. If the counter-streaming ion beams were absent, the growth rate would be adjusted to $\gamma_{max} \approx 0.29\omega_{pe}$ whereas if the electrons were absent, the growth rate would become $\gamma_{max} \approx 0.06\omega_{pe}$. This implies that the phase space mixing is largely controlled by the unstable distribution of counter-streaming electron beams, rather than counter-streaming ion beams. This is contrary to the implications by Farrell {\it et al.} \cite{farrell97,farrell98} who suggest the counter-streaming ion beams are the cause of the electrostatic instability in the wake.

The generation of the two-stream instability produces some waves which travel away from the wake. The number of the fastest growing two-stream wave mode (Eq.~\ref{eqn:kmax}) can be substituted in the dispersion relation (Eq.~\ref{eqn:wdr}) to obtain the frequency and thus the velocity of this wave mode.
\begin{eqnarray}
\omega_{ts} & = & \frac{\sqrt{15}\sqrt{\omega_{be}^2 + \omega_{bi}^2}}{2} \\
\frac{\omega_{ts}}{k} & = & \sqrt{5}V_0
\end{eqnarray}

\begin{figure}
\begin{center}
\includegraphics[width=8.5cm]{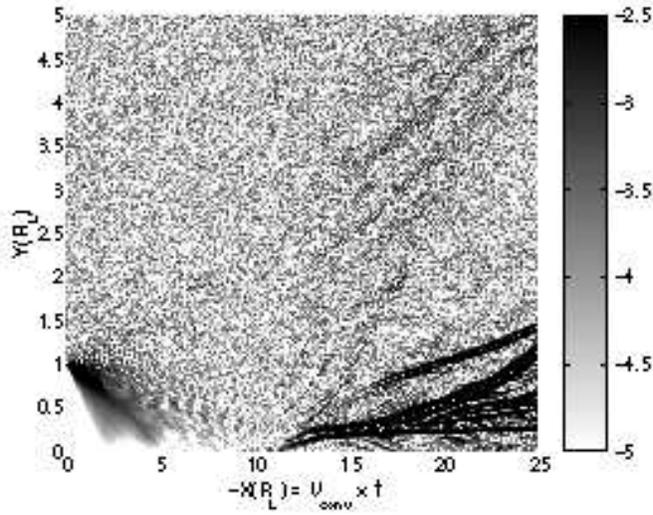}
\caption{Electric field evolution on a log scale ($\text{Vm}^{-1}$) shows distance along the simulation box Y($R_L$) against time (equated to distance behind the moon) X($R_L$). The two black lines represent a point along Y, beginning from where the electron two-stream instability begins to grow, travelling at velocities $4.2\times10^6\text{ms}^{-1}$ and $2.0\times10^6\text{ms}^{-1}$.}
\label{fig:reallogef}
\end{center}
\end{figure}

By estimating the velocity of the counter-streaming electron beams, an estimate for the velocity of these wave can be obtained. Looking at the electron phase space plots, $V_0 \approx 2v_{the} \approx 1.8\times10^6\text{ms}^{-1} \Rightarrow \omega_{ts}/k \approx 4\times10^6\text{ms}^{-1}$. Figure~\ref{fig:reallogef} shows a region outside the wake where slight increases in electric field strength were \linebreak found. The electric field magnitude is plotted on a log scale to highlight these fluctuations. The two black lines begin approximately where the two-stream instability begins to develop, and travel along Y at velocities $4.2\times10^6\text{ms}^{-1}$ and $2.0\times10^6\text{ms}^{-1}$. Notice the electric field fluctuations begin along the fastest black line. This implies that the fluctuations are in some way related to the growth of the electron two-stream instability in the wake. There is more structure in these fluctuations than can be attributed to the two-stream instability, and this matter shall be revisited in Sec.~III-F.

\subsection{Electron Phase Space Holes}

\begin{figure}
\begin{center}
\includegraphics[width=8.5cm]{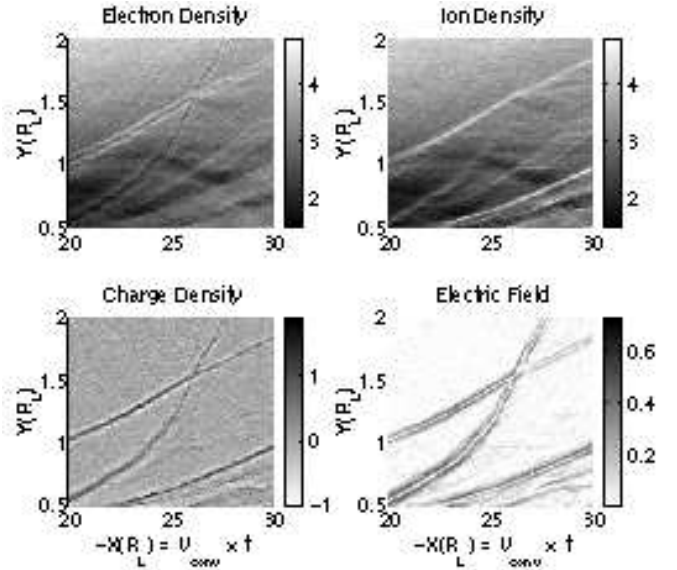}
\caption{Electron density, ion density, charge density and electric field magnitude showing two electron phase space holes passing through each other.}
\label{fig:realepsh}
\end{center}
\end{figure}

\begin{figure}
\begin{center}
\includegraphics[width=8.5cm]{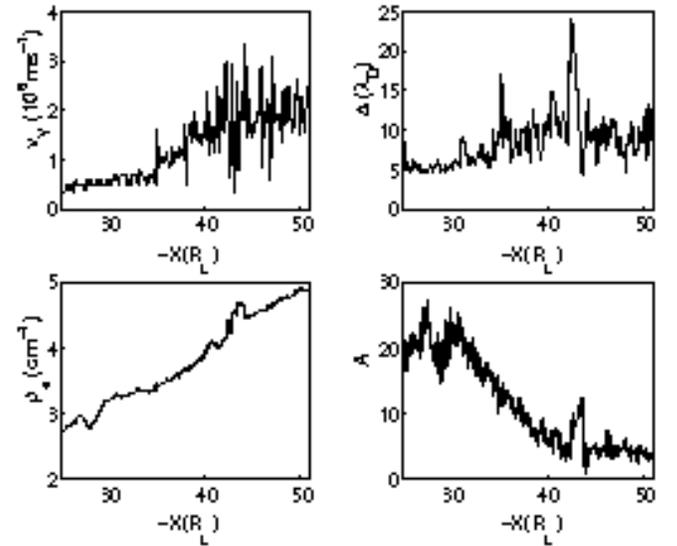}
\caption{Characteristics of an electron phase space hole, plotted against time (equated to a distance behind the Moon) X($R_L$). Shown here are 
(a) velocity along simulation box, $v_{\text{Y}}$ ($10^6\text{ms}^{-1}$),
(b) width, $\Delta$ ($\lambda_D$), 
(c) density about electron hole, $\rho_e$ ($\text{cm}^{-3}$), 
(d) amplitude, A (V).}
\label{fig:realsls31}
\end{center}
\end{figure}

As described earlier (see Figs.~\ref{fig:big}(a) and (b)), double peaked electron enhancements about a single ion peak are seen to travel away from the wake at a variety of velocities. This deviation from quasineutrality is revealed in the charge density plot (Fig.~\ref{fig:big}(c)) and can also be detected in the electric field plot (Fig.~\ref{fig:big}(d)). Figure~\ref{fig:realepsh} shows a small region of the density and electric field plots where two electron phase space holes with different velocities pass almost unaltered through each other.
By following the trajectory of the electron holes, their characteristics can be measured directly from the simulation data. The width is the distance between the peaks in the electric field magnitude. The density is the average electron density in a section of fixed length, about the centre of the electron hole. The amplitude is the height of the electric potential peak formed by the electron hole. Figure~\ref{fig:realsls31} shows the characteristics plotted against time.

The velocity of this electron phase space hole \linebreak increases as the simulation evolves. The method of measuring the velocity results in quite a noisy time series. Note the particularly noisy region at about $\text{X} = -43R_L$. At this time, the electron hole caught up with and passed another electron hole. The width of this electron phase space hole remains almost constant (disregarding the noise and the interaction with the other electron hole). The density around the electron hole, however, clearly increases with time. The electron hole passes the rarefaction wave at about $\text{X} = -40R_L$. Since the ambient electron solar wind density is at $5\text{cm}^{-3}$, the density around the electron hole should level off there. This does happen in some of the other electron holes, but the one under analysis just reaches the ambient solar wind when the simulation is stopped. The amplitude of the electron phase space hole clearly decreases as the simulation evolves. At $\text{X} = -43R_L$, when the two electron holes pass, the amplitude here will be the sum of the amplitudes. The top of the peak in amplitude, is at the time when the electron holes coincide.

\begin{figure}
\begin{center}
\includegraphics[width=8.5cm]{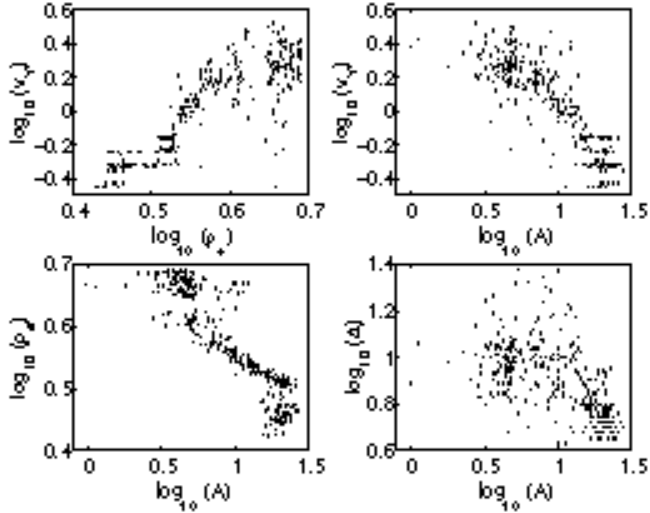}
\caption{Characteristics of an electron phase space hole, plotted against each other on log-log scales. Shown here are 
(a) velocity along simulation box, $\log_{10}(v_{\text{Y}})$ against density about electron hole, $\log_{10}(\rho_e)$,
(b) velocity along simulation box, $\log_{10}(v_{\text{Y}})$ against amplitude, $\log_{10}(\text{A})$,
(c) density about electron hole, $\log_{10}(\rho_e)$ against amplitude, $\log_{10}(\text{A})$,
(d) width, $\log_{10}(\Delta)$ against amplitude, $\log_{10}(\text{A})$.}
\label{fig:realsls33}
\end{center}
\end{figure}

To determine the relationships between these characteristics of the electron phase space hole, they are plotted against each other on a log-log plot (Fig.~\ref{fig:realsls33}).
Many of the points are due to the interaction with the other electron hole, but by considering Fig.~\ref{fig:realsls31}, the approximate positions of these can be determined. 
By disregarding these points, definite relationships between the characteristics can be found.

From Fig.~\ref{fig:realsls33}a, $v_{\text{Y}}\propto\rho_e^a$ where $a \approx 2.9$. The other electron phase space holes exhibit similar velocity - density relationships, giving $v_{\text{Y}}\tilde{\propto}\rho_e^3$. All of the other electron phase space holes exhibit various different relationships for the other characteristics.
Figure~\ref{fig:realsls33}b (velocity-amplitude plot) displays a similar structure to the theoretical and simulated electron phase space holes by Saeki and Gemna \cite{saeki98} (described later).
The structure of Fig.~\ref{fig:realsls33}c shows a region of low density, occurring at the beginning of the electron phase space hole's evolution. The more densely populated region above $\log_{10}(\rho_e) \approx 0.65$ occurs as the electron hole reaches the ambient solar wind. With other electron holes that spend longer in the ambient solar wind, this stretches out at constant density as their amplitudes decay.

\begin{figure}
\begin{center}
\includegraphics[width=8.5cm]{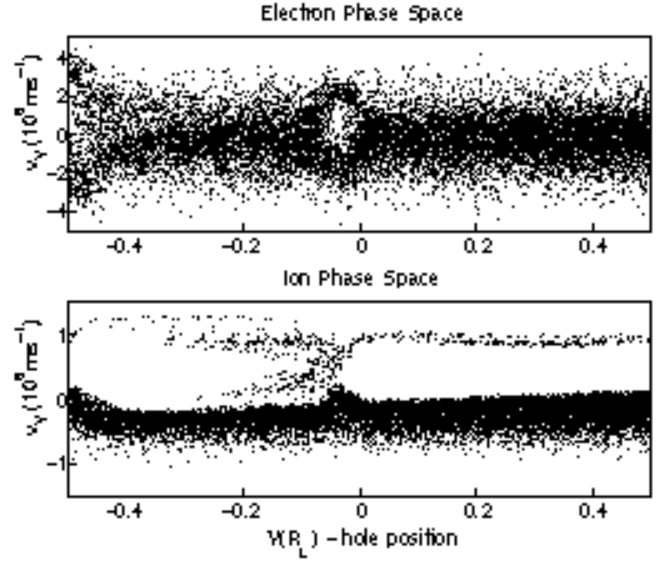}
\caption{Electron and ion phase space [$v_{\text{Y}} (10^6\text{ms}^{-1})$ against Y($R_L$)] of an electron phase space hole. In order to see the structure, a number of phase space plots over a short time have been combined with the electron hole in the same place.}
\label{fig:realsls3ps}
\end{center}
\end{figure}

The phase space structure of these electron phase space holes are shown in Fig.~\ref{fig:realsls3ps}. Notice the distinct lack of low energy electrons in the hole. The electrons around the hole are trapped in the potential peak and are ocsillating back and forth (circling in phase space). The double peaked electron density structure is produced since the electron phase space density is almost constant around the hole, thus resulting in the peaks coinciding with the hole edges and the density falling significantly in the centre. There is another larger electron phase space hole, just on the edge of this plot. The ion phase space shows a group of ions removed from the bulk of the ions, coinciding with the centre of the electron phase space hole. This is the electron hole's associated ion-acoustic soliton. The beam at the top of the ion phase space (at $v_{\text{Y}} \simeq 10^6 \text{ms}^{-1}$) does not perturb the local fields or plasma since its contribution to the total charge and current density is small, however, some of these ions are being reflected by the positive potential formed by the electron hole.

Many simulations and theoretical models have examined electron phase space holes \cite{lynov85,saeki98} with which the simulated electron holes can be compared. Saeki and Gemna \cite{saeki98} simulated an electron hole with a velocity less than the ion sound speed. There was a background of ions, which interacted with the electron hole, disrupting it in two and generating coupled states of electron holes and ion-acoustic solitons. The electron holes with a velocity above the ion sound speed do not interact in the same way with the background ions. In the lunar wake simulations, the electron phase space holes are travelling at velocities below the ion sound speed, which is where the holes are coupled with the ion-acoustic solitons. Later on, the velocity has increased above the ion sound speed and therefore the electron holes are not disrupted further but continue moving, coupled to the ion-acoustic solitons, just as in Ref.~15. Different size holes were considered and theoretical relationships between amplitude and velocity obtained. Measuring the hole phase space area from Fig.~\ref{fig:realsls3ps} gives a size, $S \approx 10\lambda_Dv_{the}$. The velocity at this time is $\approx 10^6\text{ms}^{-1}$, which gives a Mach number, $M = 4$. These measurements give a theoretical maximum potential of about 13V which is close to the measured amplitude. This implies that this model agrees well with the lunar wake simulations, indicating that these are examples of coupled states of electron holes and ion-acoustic solitons. Since a mass ratio of $m_i/m_e = 20$ has been used in the simulations, it is not clear whether such coupled states will be found in reality.

\begin{figure}
\begin{center}
\includegraphics[width=8.5cm]{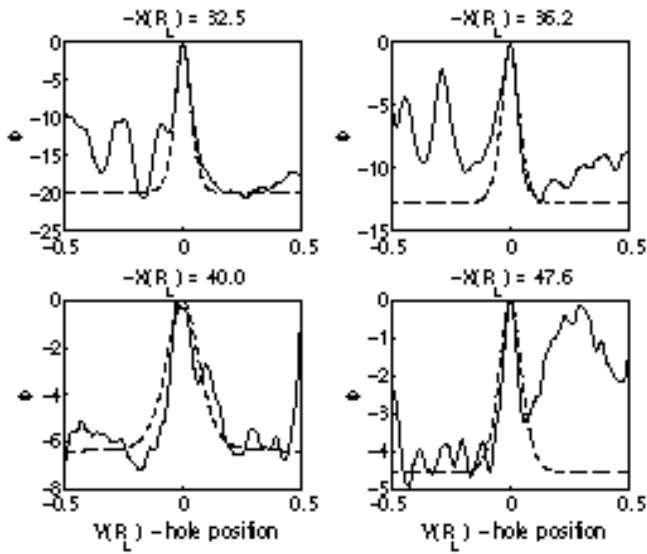}
\caption{Electric potential about electron phase space hole structure (solid line), compared to theoretical shape (dashed line)}
\label{fig:realsls37}
\end{center}
\end{figure}

Lynov {\it et al.} \cite{lynov85} used a slightly modified Bernstein-Greene-Kruskal (BGK) scheme which assumed the following form of the potential
\begin{equation}
\Phi(X) = \Phi_0{\text{sech}}^2(2X/L)
\end{equation}
where the full width at half amplitude, $\Delta X = \ln(1+\sqrt{2})L$. Figure~\ref{fig:realsls37} shows good agreement between the simulated potentials (solid line) compared to their theoretical shape (dashed line) (given the amplitude and width) for four different times. The only disagreement is in the width of the potential peak. This is due to the measurement of the widths being the distance between the two electric field strength peaks, rather than full width at half maximum. Other BGK solutions (e.g. $\Phi(X) = \Phi_0 \text{sech}^4 (2X/L)$, $\Phi(X) = \Phi_0 \text{exp}(-(2X/L)^2)$ and $\Phi(X) = \Phi_0 \text{exp}(-(2X/L)^4)$) yield slightly different forms of the potential, which still closely fit the simulation results.

\subsection{Bump-on-Tail Instability Outside  \\ the Wake}

\begin{figure}
\begin{center}
\includegraphics[width=8.5cm]{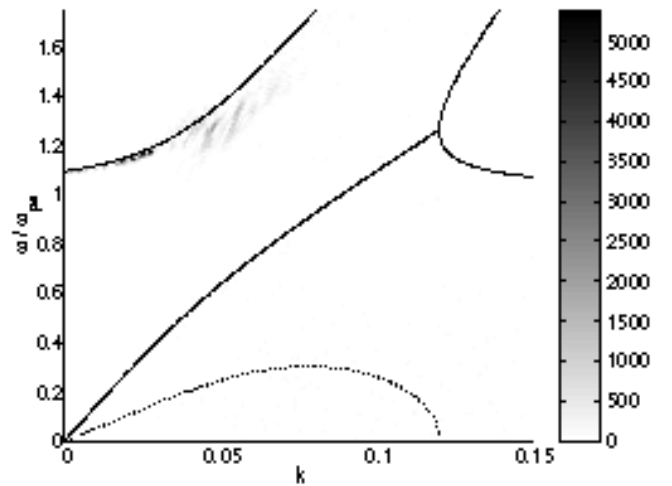}
\caption{Dispersion relation of an electron bump-on-tail instability with an ion background showing real (solid) and imaginary (dotted) parts ($V_b \approx 2\times10^6\text{ms}^{-1}$, $\omega_b \approx \omega_{pe} / \sqrt{7}$), plotted over the dispersion relation produced from a region of the simulation outside the wake where the waves occur. The frequencies have been normalised to the electron plasma frequency ($\omega_{pe}$)}
\label{fig:drebot}
\end{center}
\end{figure}

The electrons that pass right through the wake and emerge on the other side, generate a bump-on-tail distribution which is unstable. The resulting bump-on-tail distribution cannot be formed from the sum of two Maxwellians since some of the bulk electrons have been moving into the void. On the beam side of the distribution, the core Maxwellian has been truncated (see Fig.~\ref{fig:realeps}). Since the core and beam populations do not overlap, the appropriate dispersion relation is that obtained from the cold plasma approximation.
\begin{equation}
\frac{\omega_{pe}^2}{\omega^2} + \frac{\omega_{pi}^2}{\omega^2} + \frac{\omega_b^2}{(\omega - kV_b)^2} = 1
\end{equation}
where $\omega_b$ is the beam plasma frequency and $V_b$ is the beam velocity. Estimates of $\omega_b \approx \omega_{pe} / \sqrt{7}$ and $V_b \approx 2\times10^6\text{ms}^{-1}$ yields the dispersion relation shown in Fig.~\ref{fig:drebot}. The solid line represents real values of the frequency and the dotted line represents the imaginary part, or the growth rate.

Taking a region of the electric field where the waves occur, but away from any other disturbances, the dispersion relation can be compared with the theoretical one in Fig.~\ref{fig:drebot}. The wavelengths of the fluctuations in the electron density are  $\lambda \approx 100\text{m}$, which equates to a wavenumber $k \approx 0.063\text{m}^{-1}$. These waves are purely electrostatic and can only be easily seen in a section of the simulation, related to the two-stream instability growth in the wake as described in Sec.~III-D. This does not mean that the waves are only present in this section, just that they are amplified there. It is possible to see very slight fluctuations outside of this section in some electron density plots.

Finally, the simulations presented in Ref.~13 were found to have a scaling error, outlined in Ref.~14. The ions had been given a charge $\sim4.5$ times greater than they should have. This resulted in extra structures forming as described in Ref.~14. All the results presented here are from the corrected simulations.

\section{Conclusion}

Results from high resolution simulations of the lunar wake have been presented here, fully resolving the electron phase space dynamics and leading to a more complete understanding of the ion phase space dynamics. Analysis of the two-stream instability generated by the counter-streaming particle beams from either side of the wake, revealed that the instability is largely due to the presence of the counter-streaming electrons, rather than the counter-streaming ions, as previously thought. 
By examining the particle densities in detail, peaked structures associated with the propagating electron phase space holes, generated by the two-stream instability, \linebreak have been found. These structures are also evident in the electric field and electric potential plots. Characteristics of these electron phase space holes have been examined, and found to be in good comparison with previous theoretical and computational work. An electron bump-on-tail instability has been found just outside the wake, where electrons had passed through the wake and created an unstable distribution on the other side.

The variations in the IMF are expected to cause significant differences in the structure of the lunar wake. Since the IMF was at an oblique angle and varying in magnitude throughout the WIND flyby, comparisons with geometrically simple simulations such as this one are difficult. Therefore, in order to simulate these conditions more closely, higher dimensional simulations are being undertaken. Comparisons of the simulations have been made and are similar to those in Ref.~13. The main difference is that a constant electron temperature was simulated.

\section{Acknowledgments}
This work has been funded by the Particle Physics and Astronomy Research Council (PPARC) and Higher Education Funding Council for England (HEFCE) and included usage of the GRAND and Computer Services for Academic Research (CSAR) T3E facilities.

{}

\end{document}